\documentclass[a4paper,10pt,twoside]{cpc-hepnp}

\usepackage{multicol}
\usepackage{graphicx}
\usepackage{booktabs}
\usepackage{amssymb,bm,mathrsfs,bbm,amscd}
\usepackage[tbtags]{amsmath}
\usepackage{lastpage}
\usepackage{epstopdf}

\begin{document}

\fancyhead[c]{\small Submitted to \textquotedblleft Chinese Physics C\textquotedblright } \fancyfoot[C]{\small \thepage}

\title{Central frequency measurement of the HLS-II storage ring\thanks{Supported by the upgrade project of Hefei light source}}

\author{%
      ZHENG Jia-Jun$^{1;1)}$\email{zhengjj@mail.ustc.edu.cn}%
\quad YANG Yong-Liang$^{1;2)}$\email{ylyang@ustc.edu.cn}%
\quad SUN Bao-Gen$^{1}$\\
\quad WU Fang-Fang$^{1}$
\quad CHENG Chao-Cai$^{1}$
\quad TANG Kai$^{1}$
}
\maketitle

\address{%
$^1$ National Synchrotron Radiation Laboratory, University of Science and Technology of China, Hefei 230029, China
}

\begin{abstract}
Central frequency is a key parameter of storage rings. This paper presents the measurement of central frequency of the HLS-II storage ring using the sextupole modulation method. Firstly, the basis of central frequency measurement of the electron storage ring is briefly introduced. Then, the error sources and the optimized measurement method for the HLS-II storage ring are discussed. The workflow of the self-compiled Matlab script used in central frequency measurement is also described. In the end, the results achieved by using two methods to cross-check each other are shown. The measured value of the central frequency demonstrates that the real circumference of the HLS-II storage ring agrees well with the designed value.
\end{abstract}

\begin{keyword}
HLS-II storage ring, circumference, central frequency, sextupole modulation
\end{keyword}

\begin{pacs}
29.20.db, 29.85.Ca, 29.90.+r
\end{pacs}

\footnotetext[0]{\hspace*{-3mm}\raisebox{0.3ex}{$\scriptstyle\copyright$}2013
Chinese Physical Society and the Institute of High Energy Physics
of the Chinese Academy of Sciences and the Institute
of Modern Physics of the Chinese Academy of Sciences and IOP Publishing Ltd}%

\begin{multicols}{2}

\section{Introduction}

The upgraded Hefei light source (HLS-II) at National Synchrotron Radiation Laboratory (NSRL) is a storage-ring-based light source which provides the scientific users high brightness photons with an energy range from VUV to far IR. It has been well operated since the commissioning was completed in the late 2014~\cite{bib:1}. The upgrade of the 800MeV storage ring is the major part of the upgrade project. On one hand, the lattice is replaced by DBA (double bend achromatic) cells~\cite{bib:2}, which not only significantly decreases the beam emittance but also increases the straight sections for insertion devices. On the other hand, a series of subsystems, such as RF system, power supply system and corresponding instrumentations, have been redeveloped and rebuilt in order to improve the machine stability and reliability. The main parameters of HLS-II storage ring are shown in Tab.~\ref{tab:1}.

The central frequency is a direct measure of the circumference of the storage ring. The central frequency can be used to estimate the deviation of the ring circumference from the designed value after the ring installation. In addition, the evolution of the circumference induced by various effects such as tide need be tracked in long term using the beam position measurement. But the average radical beam position should be calibrated by the central frequency measurement~\cite{bib:3}.

Moreover, the beam passing off-center through the quadrupole can \textquotedblleft sense\textquotedblright the coexistence of the dipole and quadrupole fields in a separated function lattice, which changes the distribution between the horizontal and the longitudinal partition numbers. By increasing the RF frequency, which in essence increases the horizontal partition number, the emittance of the torage ring can be further reduced~\cite{bib:4}. However, this adjustment is limited by the instability of the longitudinal motion since the longitudinal partition number is reduced at the same time. Therefore, the adjustable range of the RF frequency in this case must be investigated by comparing the current RF frequency with the central frequency.
\begin{center}
\tabcaption{ \label{tab:1}  Main parameters of HLS-II storage ring.}
\footnotesize
\begin{tabular*}{80mm}{c@{\extracolsep{\fill}}c}
\toprule Parameter & Value    \\
\hline
Energy & 800 MeV  \\
Circumference & 66.13 m  \\
Beam current & 300 mA \\
DBA cells&4\\
RF frequency & 204.030 MHz  \\
Harmonic number & 45 \\
Tune & 0.44/0.36 \\
Emittance & 36 nm.rad \\
\bottomrule
\end{tabular*}
\end{center}

\section{Basis of the central frequency measurement}

The beam orbit in the storage ring is defined by the arrangement of bending magnets. The quadrupoles for focusing and sextupoles for chromaticity and nonlinearity correcting are distributed along the orbit and aligned to make sure that the beam orbit passes the magnetic centers. And the orbit passing on average through the center of the quadrupoles and sextupoles is called the central orbit and the length of the central orbit is the circumference of the ring. Therefore, the RF frequency corresponding to the central orbit is called the central frequency. Using the relation between the RF frequency and the length of the beam orbit for relativistic particles, the central RF frequency $f_{RF}^{C}$ is defined as
\begin{equation}
f_{RF}^{C}=h\frac{c}{C_{C}},
\label{equ:1}
\end{equation}
where $h$ is the harmonic number, $C_{C}$ is the length of the central orbit and $c$ is the speed of light. However, magnets alignment errors cannot be avoided in the practical installing of the machine. There is always slight difference between the practical circumference and the designed value of the storage ring.

The RF frequency differing from the central frequency leads to two effects. On one hand, the beam will \textquotedblleft sense\textquotedblright coexistence of dipole and quadrupole fields in the location of quadrupole magnets. The distribution of the damping partition numbers in three degrees of freedom can be modified  according to Robinson¡¯s criterion~\cite{bib:5}. Therefore, the central RF frequency can be derived by the relation between the RF frequency and either damping partition numbers or other dependent parameters~\cite{bib:6}. However, the accuracy of this method is seriously limited due to the strong collective effects induced by the high beam current. These methods have not been considered yet until the instability issues in HLS-II storage ring are systematically studied.

On the other hand, the beam will also \textquotedblleft sense\textquotedblright extra quadrupole fields in the location of sextupoles. The method basing on this effect refers as to the direct measurement of central frequency which is a standard methods with both simplicity and reliability~\cite{bib:7}.

If the current RF frequency differs from the central RF frequency, the relative momentum deviation $\delta$ is

\begin{equation}
\delta=-\frac{1}{\alpha}\frac{\Delta f_{RF}}{f_{RF}^{C}},
\label{equ:2}
\end{equation}
where $\alpha$ is the momentum compact factor and $\Delta f_{RF}$ is the deviation of current RF frequency from the central frequency. Then the off-momentum particles will "sense" an extra focusing force introduced by sextupole fields in the dispersion section. Consequently, the tune shift is
\begin{equation}
\Delta\upsilon_{x,y}=\delta\xi_{x,y},
\label{equ:3}
\end{equation}
where $\xi_{x,y}$ is the corrected chromaticity. And the corrected chromaticity is relative to the sextupole setting as
\begin{equation}
\xi_{x,y}=-\frac{1}{4\pi}\oint\beta_{x,y}[K_{x,y}(s)-S(s)\eta(s)]ds,
\label{equ:4}
\end{equation}
where $S$ is the sextupole strength, $\beta_{x,y}$ is the beta function, $\eta$ is the dispersion function, $K_{x,y}$ is the quadrupole strengths and the subscripts represent the two transverse directions.

From Eq.~\ref{equ:2}, \ref{equ:3} and \ref{equ:4}, it is clear that the tune will not vary with the different sextupole settings if $\Delta f_{RF}$ is equal to zero. Therefore, the measurement of central frequency of a storage ring is to find the RF frequency points at which the chromaticity curves cross, or in other words, to find the frequency point at which the tune does not rely on the sextupole strength. The methods to find this RF frequency point can be classified into two categories: RF shaking and sextupole modulation~\cite{bib:8}. In RF shaking method, the RF frequency is repeatedly swept as a predefined function for the different chromaticity settings. Then the frequency at the crossing point of the different chromaticity curves is viewed as the central frequency. In opposite to RF shaking, sextupole modulation is to sweep repeatedly the strength of the sextupoles at different RF frequency. Although the two methods to obtain the data differ, there is no essential distinction between them.
\section{Error source analysis and measurement optimizing}

In practical, the central frequency achieved by direct measurement is a measure of the length of the central orbit in the sextupoles. The quadrupoles in the storage ring are not concerned. Thus the accuracy of central frequency measurement is limited by the alignment errors between sextupoles and quadrupoles. But this effect is usually small. Fig.~\ref{fig:1} shows the adjacent quadrupoles and sextupoles in a DBA cell of the HLS-II storage ring. Note that the four magnets are installed on the same port as a unit. And the adjacent sextupoles and quadrupoles are aligned with respect to each other with the accuracy lower than 0.08 mm ~\cite{bib:9}. On the other hand, the HLS-II storage ring uses two sextupole families (denote as S3 and S4 and each family has 8 identical sextupoles) in the dispersion sections to correct the chromaticity~\cite{bib:10}. And the Eq.~\ref{equ:4} becomes
\begin{equation}
\xi_{x}=\xi_{x}^{nat}+\frac{2}{\pi}(s_{3}\eta_{3}\beta_{x3}l_{3}+s_{4}\eta_{4}\beta_{x4}l_{4}),
\label{equ:5}
\end{equation}
\begin{equation}
\xi_{y}=\xi_{y}^{nat}+\frac{2}{\pi}(s_{3}\eta_{3}\beta_{y3}l_{3}+s_{4}\eta_{4}\beta_{y4}l_{4}),
\label{equ:6}
\end{equation}
where $\xi_{x}^{nat}$  and $\xi_{y}^{nat}$ are the natural chromaticity in two directions which are sextupoles-independent, $s_{3}$ and $l_{3}$ are respectively the strength and length of the magnets in S3, while $s_{4}$ and $l_{4}$ are respectively the strength and length of the magnets in S4. The different chromaticity parameters can be obtained by modifying the strength of either S3 or S4. In order to further reduce the effect of alignment errors between magnets, measurements were independently performed respectively using S3 and S4 as reference. In addition, the measurements in two transverse directions lead to two independent results. Therefore, there are four independent measured results which can cross-check each other.

The measurement of central frequency also demands a stable machine condition. The different chromaticity curves are obtained by measuring the tune shift with RF frequency in different sextupole strengths settings, which is called RF shaking. But frequent RF frequency changing may affect the stability of the machine and lead to beam loss or even crash the measurement. Moreover, the time interval for the rebuilding of the RF field in the RF cavity extends the measurements time, which increases the probability of occurrence as well. In order to obtain stable and reliable results, the sextupole modulation method which is faster and more robust than RF shaking method was adopted in the central frequency measurement of HLS-II storage ring.

\begin{center}
\includegraphics[width=7cm]{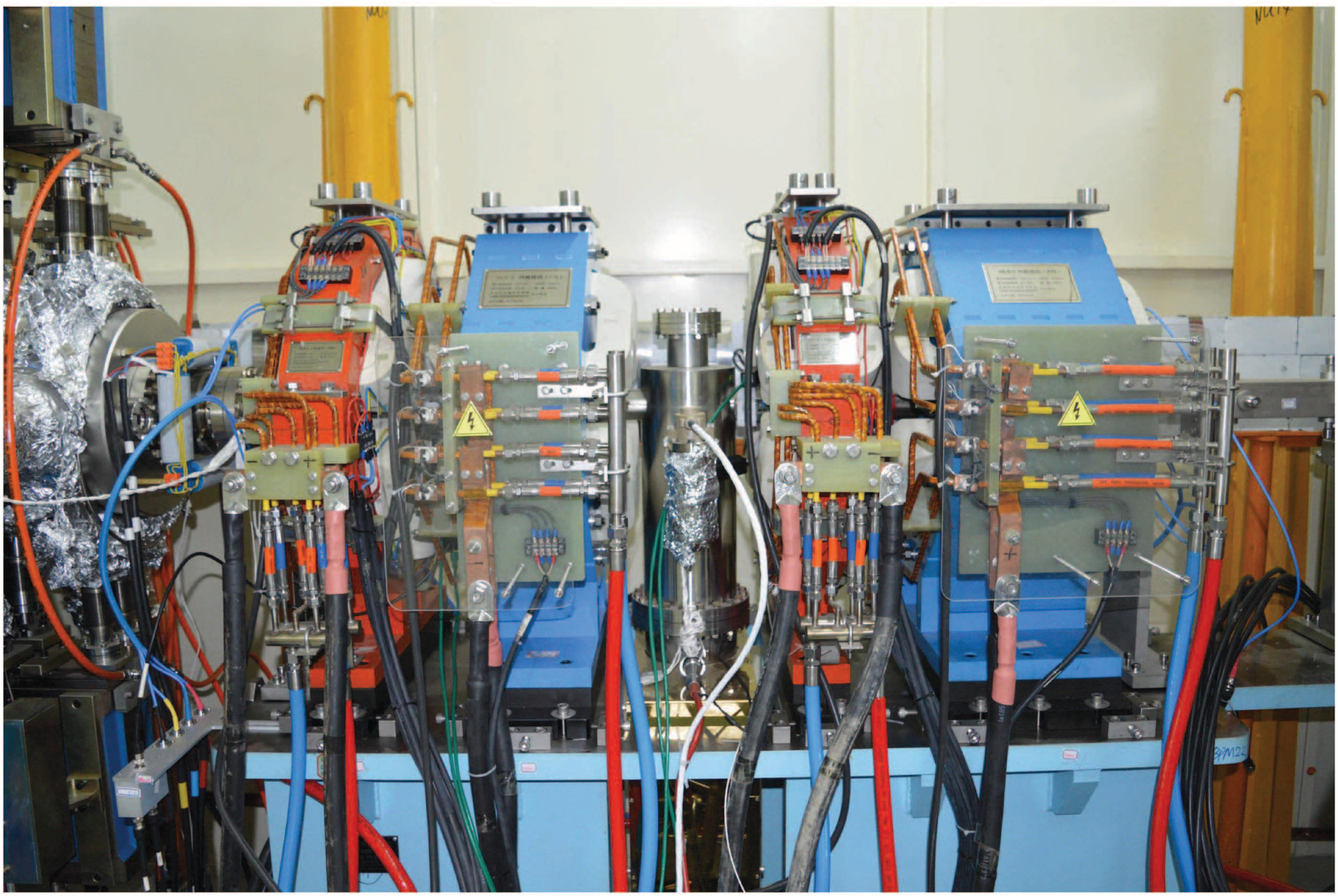}
\figcaption{\label{fig:1}Adjacent quadrupoles and
sextupoles in a DBA cell of the HLS-II storage ring (blue: Quadrupole pair, red: Sextupoles in family S3 and family S4).}
\end{center}

Another factor to affect the measurement precision is the fitting error since the central frequency point is found by fitting the measured tune shift. HLS-II storage ring is equipped with a swept-frequency based tune measurement system. The tune resolution is determined by the swept range. Resolution of 0.0001 can be achieved in the swept range from 2.1 MHz to 4.1 MHz. If the changing amount of the tune is less than or approximately equal to the resolution of the tune measurement system, the fitting error of data will seriously affect the measurement accuracy. However, large amount of tune shift may lead the working point to traveling across the resonance lines. In addition, for the practical consideration, the adjustable range of the RF system is limited by the nonlinear chromaticity effect and the RF system stability. Therefore, each tune shift need be predefined roughly with a balance between the resolution consideration and validity consideration. Using Eq.~\ref{equ:2},~\ref{equ:3},~\ref{equ:5} and~\ref{equ:6}, the proper RF frequency and sextupole strength for predefined tune shift can be estimated.

The measurement is performed by Matlab script under the epics frame. The data interaction are based on MCA (Matlab Channel Access) toolbox~\cite{bib:11} which provides interfacing between Matlab and epcis. The flow diagram of the measurement is shown in Fig.~\ref{fig:2}. The script consists of two layers of circulation to perform the sextupole modulation. The outer is the data setting and reading of RF frequency and the inner is the data setting and reading of sextupole strengths of either S3 or S4. The time interval between two data setting events is 30 seconds to make sure that the stable machine parameters are recorded into Matlab workspace. In order to monitor whether any beam loss occurred or not, the beam current is sampled when every inner circling was completed during the measurement. When the measurement is done, a preprocessing of the data is immediately made to investigate the validity of the measured data. In this processing, the beam current changing and the working point movement during the measurement can be  observed, as well as the linearity of tune shift with sextupole strength and RF frequency  .

\begin{center}
\includegraphics[width=6cm]{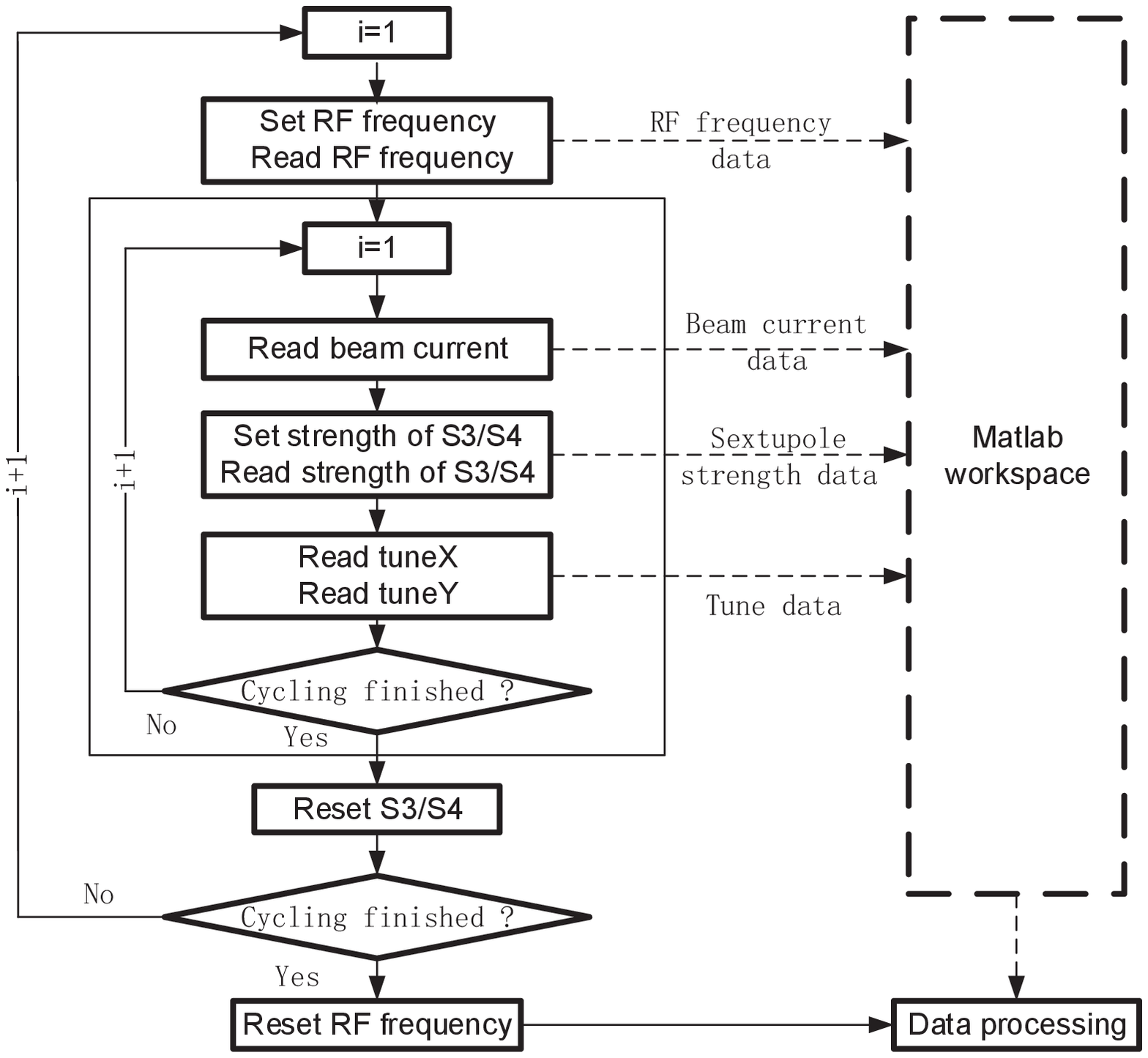}
\figcaption{\label{fig:2}The flow diagram of the central frequency measurement using S3/S4 modulation.}
\end{center}

\section{Results and discussion}

The measurements was started at the beam current of 3.62mA in the single bunch mode. The collective effects can be seen unconsidered in a large extent in this mode. The RF frequency and strength of S3 (or S4) were respectively swept 5 times as a predefined function. The sampled beam current data is shown in Fig.~\ref{fig:3}. There was no significant beam loss when the measurements were performing. The traveling of the working point during the measurement is shown in Fig.~\ref{fig:4} (with the resonance lines up to fifth order). The maximum of the working point movement are approaching to 0.02 in both directions. Still, each working point is far away from the dominant resonance lines.

\begin{center}
\includegraphics[width=7cm]{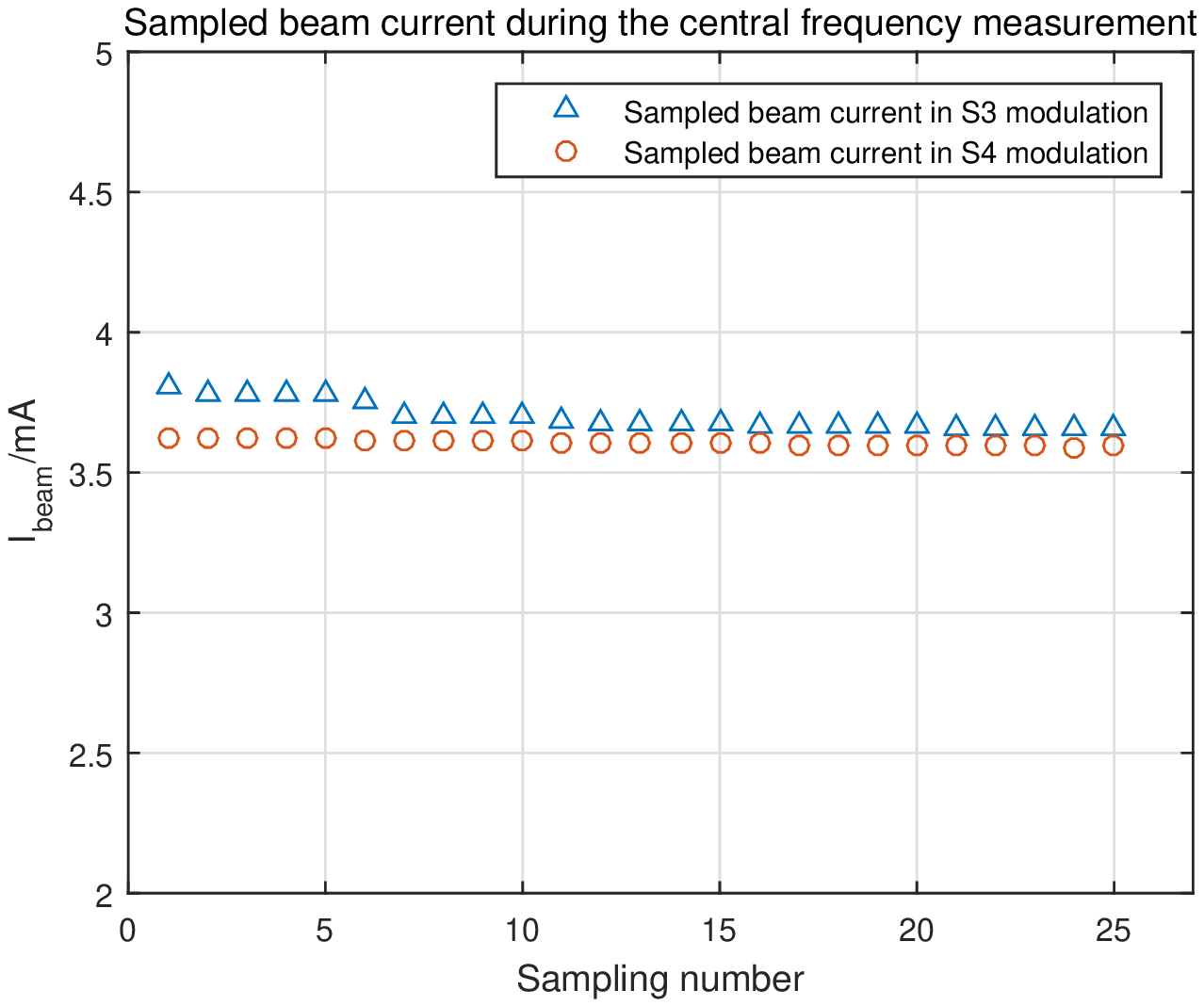}
\figcaption{\label{fig:3}The sampled beam current during the central frequency measurement.}
\end{center}

\begin{center}
\includegraphics[width=7cm]{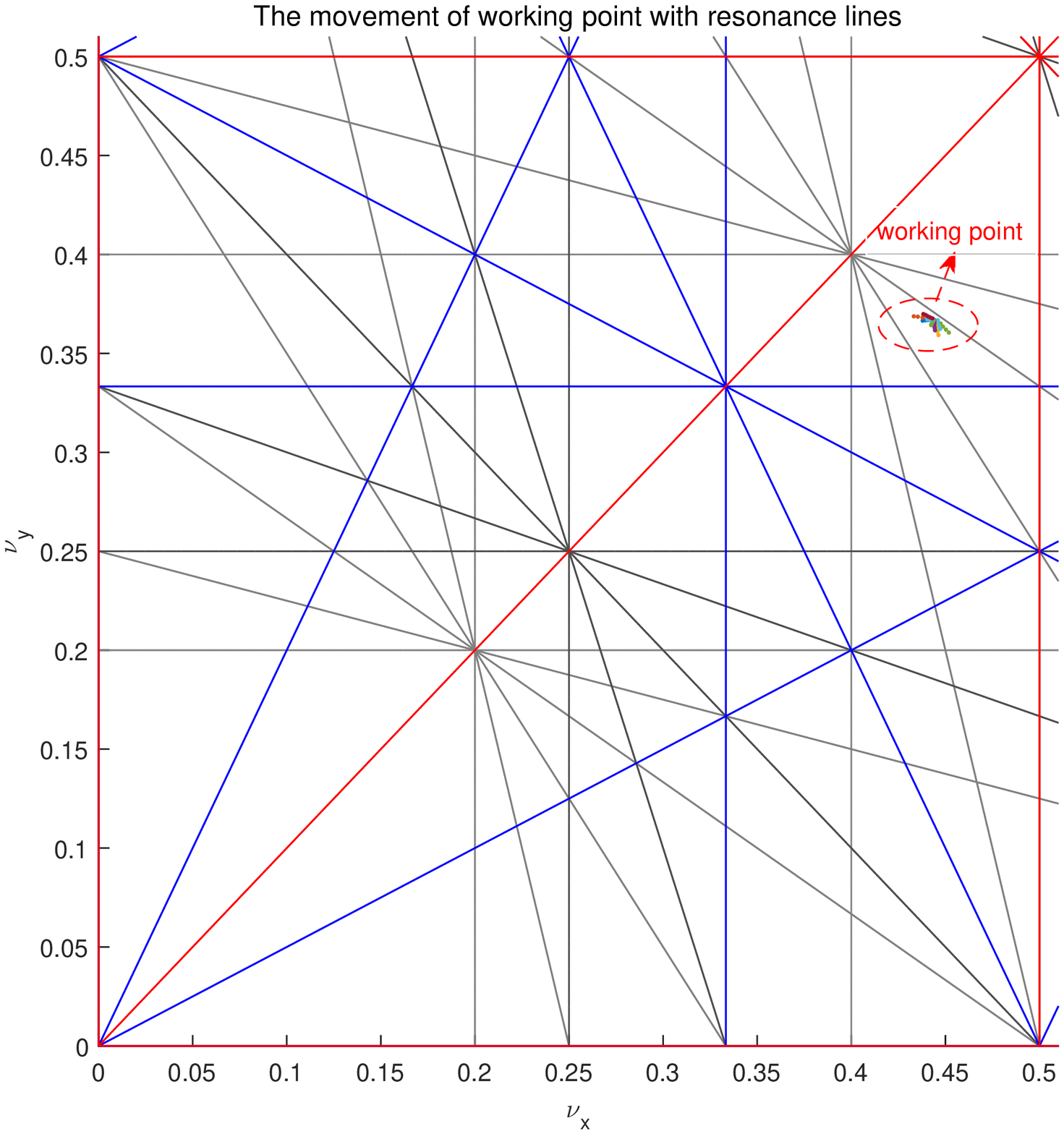}
\figcaption{\label{fig:4}The working point movement during the central frequency measurement (with the resonance lines up to fifth order).}
\end{center}

Fig.~\ref{fig:5} shows the horizontal and vertical chromaticity curves at different strengths of S3. The average frequency with corresponding rms value of the crossing points in horizontal is 204.027005$\pm$0.000762 MHz, and the obtained result in vertical is 204.027546$\pm$0.000148 MHz. Similarly, the chromaticity curves in the measurement using S4 modulation are shown in Fig.~\ref{fig:6}. The results are respectively 204.027322$\pm$0.000119 MHz and 204.027342$\pm$0.000030 MHz.

\begin{center}
\includegraphics[width=7cm]{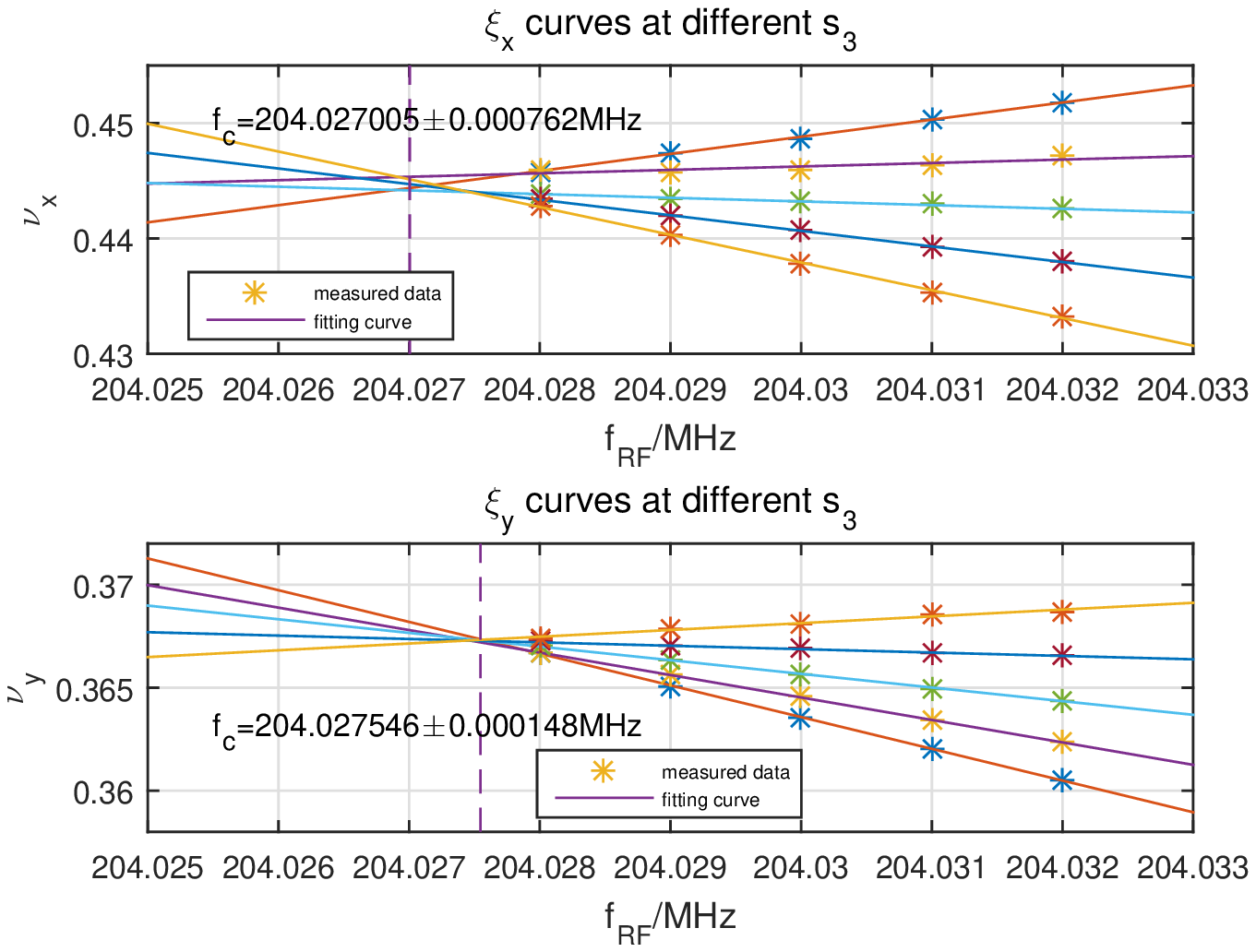}
\figcaption{\label{fig:5}Chromaticity curves at different strengths of S3.}
\end{center}

\begin{center}
\includegraphics[width=7cm]{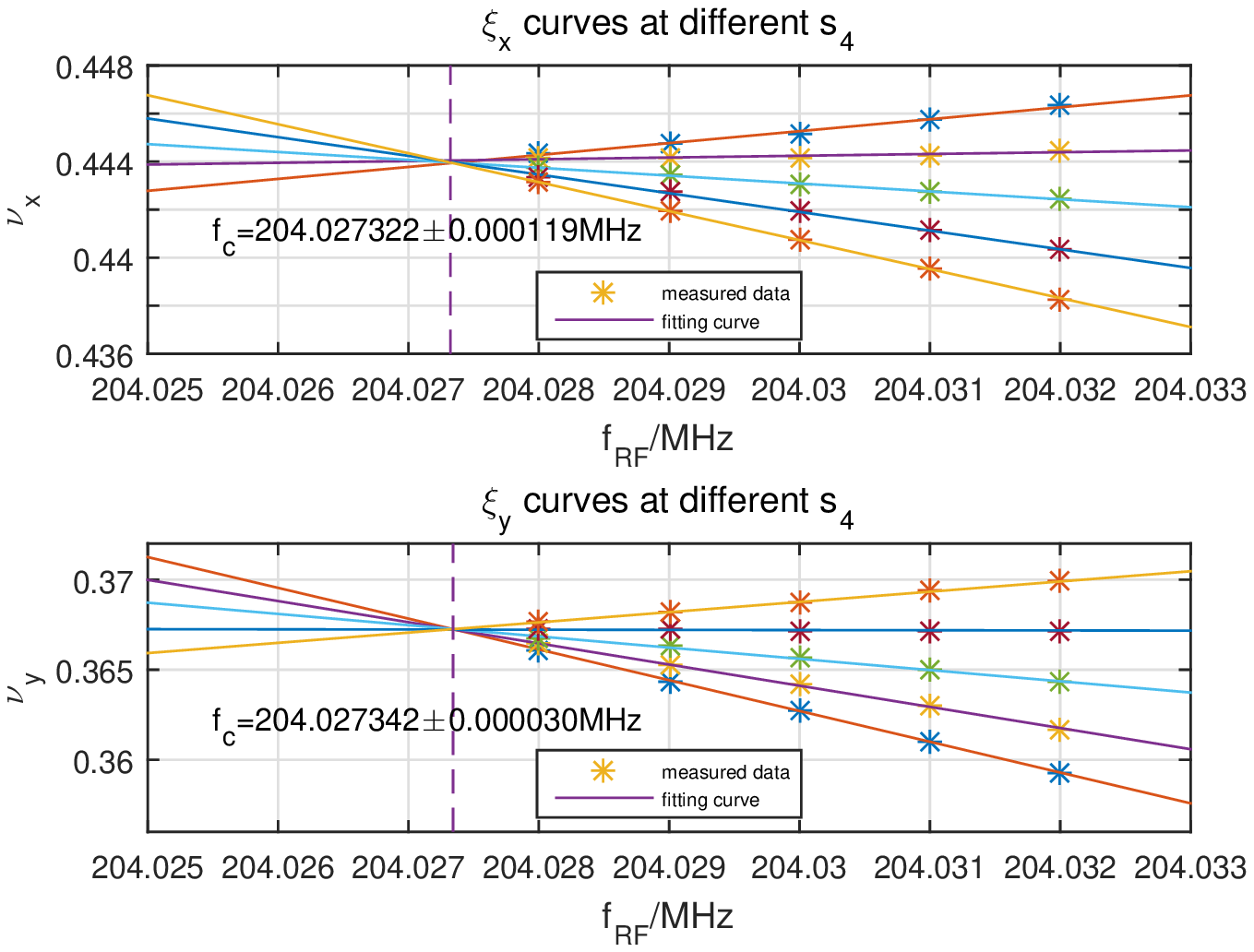}
\figcaption{\label{fig:6}Chromaticity curves at different strengths of S4.}
\end{center}

The chromaticity curves are achieved by fitting the tune shift with RF frequency. If the amount of tune shift with the RF frequency is smaller than the resolution of tune measurement system, the fitting error would be significant to find the central frequency point. To compensate this, the existing data were processed in another method. From Eq.~\ref{equ:5} and~\ref{equ:6} we can see that the chromaticity changes linearly with the strength of single sextupole family either S3 or S4. The central frequency can be obtained by finding the frequency point at which the dependence of tunes on strength of sextupole vanishes. The dependence coefficients were obtained by linear fitting of tune shift with strengths of sextupoles. Fig.~\ref{fig:7} and Fig.~\ref{fig:8} shows respectively the dependence coefficients in two transverse directions as a function of RF frequency in S3 and S4 modulations. The obtained results agree well with each other.

We take the vertical result of S4 modulation measurement as a credible value of the central frequency of the HLS-II storage ring since the uncertainty is the lowest and this value is also covered by other results.  Using the Eq.~\ref{equ:1}, the relative deviation of the circumference from the designed value after the installation of the ring is $1.3027\times10^{-5}$. And the circumference deviation is  $8.6151\times10^{-4}$ m.

\begin{center}
\includegraphics[width=7cm]{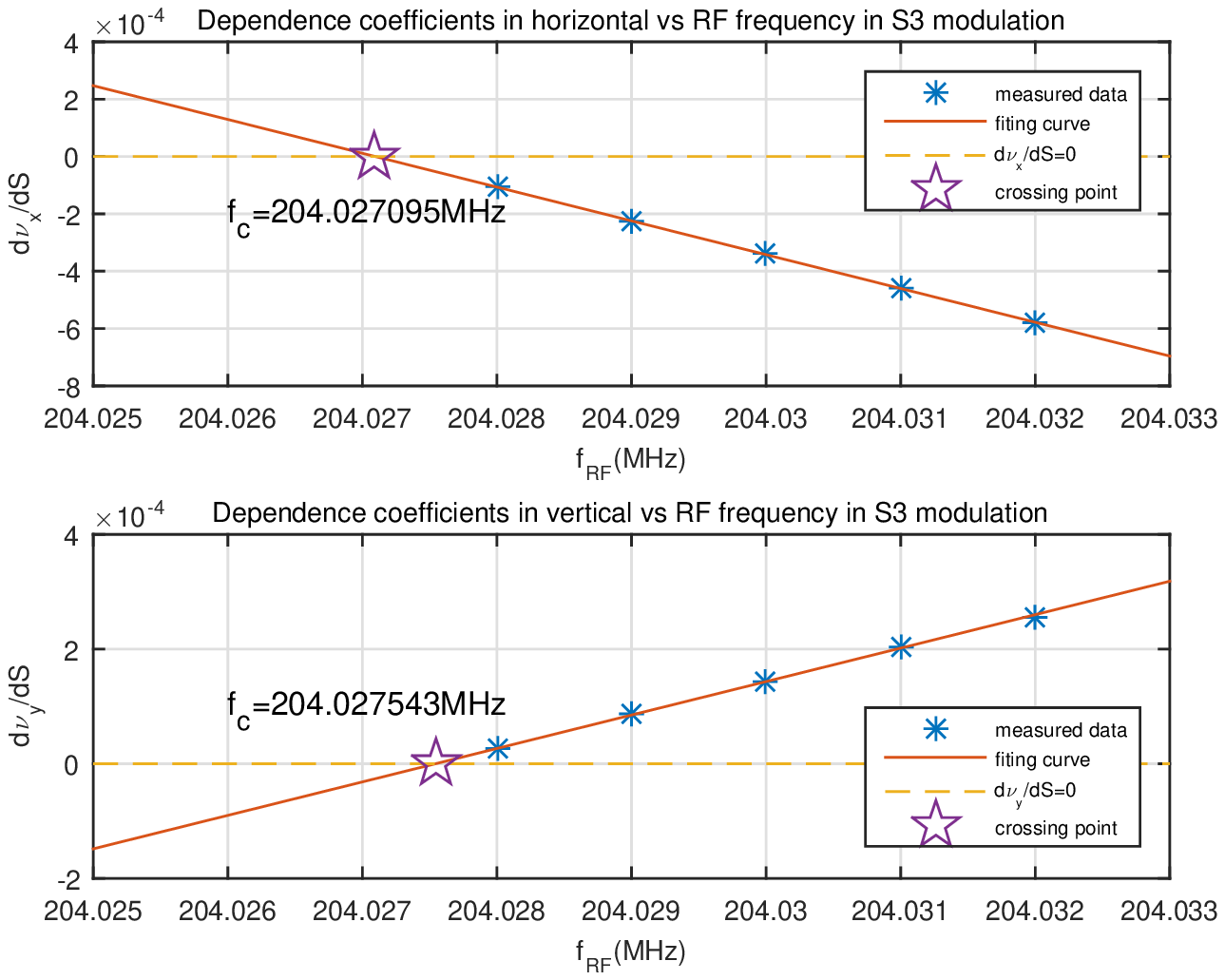}
\figcaption{\label{fig:7}The dependence coefficients of tunes on the strengths of S3 at different RF frequency.}
\end{center}

\begin{center}
\includegraphics[width=7cm]{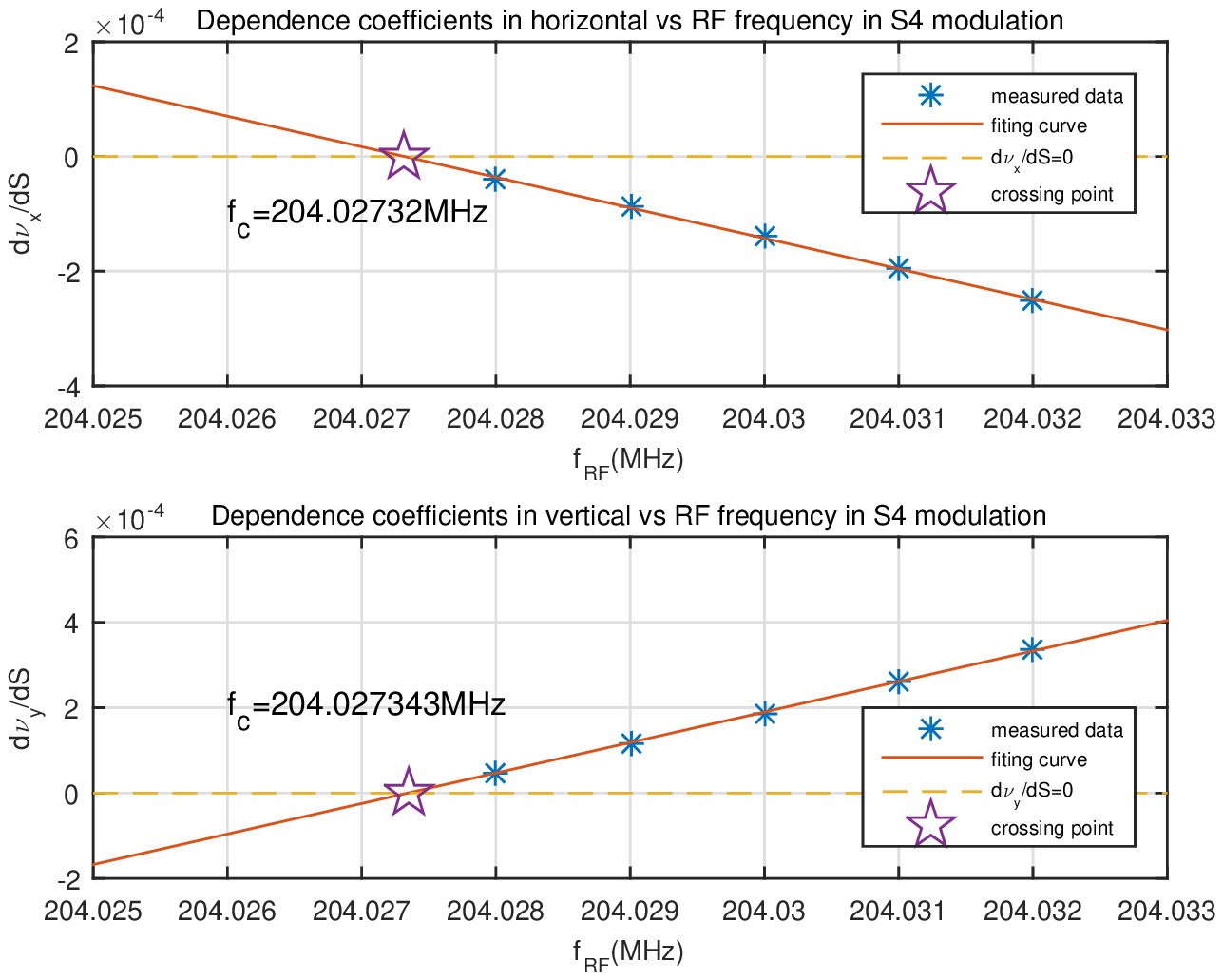}
\figcaption{\label{fig:8}The dependence coefficients of tunes on the strengths of S4 at different RF frequency.}
\end{center}

\vspace{1mm}

\section{Conclusion}

During the commissioning phase of HLS-II storage ring, the central frequency was measured using the sextupole modulation method. Three factors to affect the measurement accuracy were considered before measuring in order to optimize the measurement method. These factors include the alignment error of the magnets, the machine stability during the measurement and the fitting error resulting from the limited resolution of the tune measurement system.  Four independent groups of data were obtained by the measurements in two directions and using sextupole families S3 and S4 modulation. The amount of tune shift were optimized with the balance between the stability and resolution of the tune measurement sysytem. The measurements were performed using the self-compiled Matlab script which works under the EPICS frame. During the measurement, there is no significant beam loss. By processing the four groups of measured data in two methods, the results can be cross-checked. The results demonstrates that the circumference deviation of the HLS-II storage ring is lower than 1 mm. Moreover, the measurement result can be used for future use such as the long-term circumference tracking and the emittance optimizing of the HLS-II storage ring.

\end{multicols}

\vspace{-1mm}
\centerline{\rule{80mm}{0.1pt}}
\vspace{2mm}

\begin{multicols}{2}

\end{multicols}

\clearpage

\end{document}